\documentclass[aps,prd,twocolumn,showpacs,amsmath,amssymb]{revtex4-1}
\usepackage{amsmath}
\usepackage{graphicx}
\usepackage{subfigure}
\usepackage{epstopdf}
\usepackage{color}
\usepackage{multirow}
\usepackage{setspace}
\usepackage{overpic}
\usepackage{amssymb}
\usepackage[bookmarksnumbered, pdfstartview=FitH,colorlinks,urlcolor=blue, citecolor=blue,linkcolor=blue] {hyperref}
\usepackage{lineno}
\usepackage{bm}
\usepackage{rotating}
\usepackage[utf8]{inputenc}

\let\oldequation\equation
\let\oldendequation\endequation
\renewenvironment{equation}
  {\linenomathNonumbers\oldequation}
  {\oldendequation\endlinenomath}

\begin{document}

\title{\bf \boldmath
Search for $e^{+}e^{-} \to \phi \chi_{c0}$ and $\phi\eta_{c2}(1D)$ at center-of-mass energies from 4.47 to 4.95 GeV}

\author{
M.~Ablikim$^{1}$, M.~N.~Achasov$^{4,c}$, P.~Adlarson$^{76}$, O.~Afedulidis$^{3}$, X.~C.~Ai$^{81}$, R.~Aliberti$^{35}$, A.~Amoroso$^{75A,75C}$, Q.~An$^{72,58,a}$, Y.~Bai$^{57}$, O.~Bakina$^{36}$
% ...
\\
\vspace{0.2cm}
(BESIII Collaboration)\\
\vspace{0.2cm} {\it
$^{1}$ Institute of High Energy Physics, Beijing 100049, People's Republic of China\\
% ...
$^{3}$ Bochum  Ruhr-University, D-44780 Bochum, Germany\\
$^{4}$ Budker Institute of Nuclear Physics SB RAS (BINP), Novosibirsk 630090, Russia\\
% ...
$^{35}$ Johannes Gutenberg University of Mainz, Johann-Joachim-Becher-Weg 45, D-55099 Mainz, Germany\\
$^{36}$ Joint Institute for Nuclear Research, 141980 Dubna, Moscow region, Russia\\
% ...
$^{57}$ Southeast University, Nanjing 211100, People's Republic of China\\
$^{58}$ State Key Laboratory of Particle Detection and Electronics, Beijing 100049, Hefei 230026, People's Republic of China\\
% ...
$^{72}$ University of Science and Technology of China, Hefei 230026, People's Republic of China\\
% ...
$^{75}$ University of Turin and INFN, (A)University of Turin, I-10125, Turin, Italy; (B)University of Eastern Piedmont, I-15121, Alessandria, Italy; (C)INFN, I-10125, Turin, Italy\\
$^{76}$ Uppsala University, Box 516, SE-75120 Uppsala, Sweden\\
% ...
$^{81}$ Zhengzhou University, Zhengzhou 450001, People's Republic of China\\
\vspace{0.2cm}
$^{a}$ Deceased\\
% ...
$^{c}$ Also at the Novosibirsk State University, Novosibirsk, 630090, Russia\\
% ...
}
}

\begin{abstract}
Utilizing a data set of $6.7$ fb$^{-1}$ from electron-positron collisions recorded by the BESIII detector at the BEPCII storage ring, a search is conducted for the processes $e^{+}e^{-} \to \phi \chi_{c0}$ and $\phi\eta_{c2}(1D)$ across center-of-mass energies from 4.47 to 4.95 GeV.  In the absence of any significant signals, upper limits are set.  These include limits on the  Born cross sections for $e^{+}e^{-} \to \phi \chi_{c0}$, as well as the product of the Born cross section for $e^{+}e^{-} \to \phi\eta_{c2}(1D)$ and a sum of five branching fractions. Furthermore, the product of the electronic width of $Y(4660)$ and the branching fraction of the $Y(4660) \to \phi\chi_{c0}$, denoted as $\Gamma^{Y(4660)}_{e^{+}e^{-}} \mathcal{B}_{Y(4660) \to \phi\chi_{c0}}$, is determined to be $< 0.40$ eV at the 90\% confidence level.
\end{abstract}

\maketitle

\oddsidemargin  -0.2cm
\evensidemargin -0.2cm

\section{Introduction}
In the past two decades, a set of charmonium-like states (originally named $XYZ$ but now referred to as $T_{c\bar{c}}$ states) has been observed~\cite{Workman:2022ynf}. These discoveries have enriched the hadron spectrum significantly, and provide novel ways to improve our knowledge of non-perturbative strong interactions in the $\tau$-charm energy region~\cite{Brambilla:2010cs,Yuan:2022lxf}. Among these charmonium-like states, the vector mesons $Y$ with quantum number $J^{PC}=1^{--}$, such as the $Y(4260)$~\cite{BaBar:2005hhc,CLEO:2006ike,Belle:2007dxy} (renamed $Y(4230)$, due to the smaller mass measured by BESIII~\cite{BESIII:2016bnd}), $Y(4360)$~\cite{BaBar:2006ait,Belle:2007umv}, and $Y(4660)$~\cite{Belle:2007umv} are inconsistent with the structures observed in the line shape of the inclusive cross section of electron-positron annihilations into hadrons. Moreover, the number of these charmonium-like states exceeds the predicted number of states in the potential model~\cite{Yuan:2019zfo,BESIII:2020nme}. So they are unlikely to be conventional $c\bar{c}$ charmonium states, and are candidates for exotic states with more complex internal structures. 

The $Y(4660)$ is the heaviest vector charmonium-like state that has been well-established experimentally. It was firstly observed in the $\pi^+ \pi^- \psi(3686)$ final state by the Belle experiment~\cite{Belle:2007umv}, then confirmed later via the channels $\pi^+ \pi^- \psi(3686)$~\cite{BaBar:2006ait,BaBar:2012hpr}, $D_{s}^+D_{s1}(2536)^-$~\cite{Belle:2019qoi} and $\Lambda_c^+\Lambda_c^-$~\cite{Belle:2008xmh} in the BaBar and Belle experiments. There have been many theoretical models proposed to interpret the $Y(4660)$, such as a tetra-quark ($[cs][\bar{c}\bar{s}]$ or $[cq][\bar{c}\bar{q}]$)~\cite{Tan:2019knr,Lebed:2016yvr,Deng:2019dbg,Lu:2016cwr,Chen:2010ze,Sundu:2018toi,Chiu:2005ey,Maiani:2014aja}, conventional charmonium~\cite{Ding:2007rg,Li:2009zu}, molecules consisting of $f_{0}(980)\psi(2S)$, $\Lambda_c\bar{\Lambda}_c$, $D\bar{D}$, or a light-hadron charmonium pair~\cite{Guo:2008zg,Salnikov:2023qnn,Liu:2013rxa,Peng:2022nrj} and a $c\bar{c}g$ hybrid~\cite{Braaten:2014qka,Chen:2021aud}. The dynamical mechanisms underlying these models involve the (color-screened) potential model~\cite{Lu:2016cwr}, (quenched) lattice QCD~\cite{Chiu:2005ey,Braaten:2014qka}, kinematic effects~\cite{Ma:2014ofa}, light-cone sum rules~\cite{LCSR,Gubernari:2023rfu}, final state interaction~\cite{Chen:2023def} and mass-difference analogies~\cite{Zhu:xyzstates}. But none of them is widely accepted yet by the high energy physics community. More studies from both theoretical and experimental sides are highly desired.   

Recently, the BESIII collaboration measured a series of cross sections of electron-positron annihilation into exclusive final states, e.g., $\pi^+ \pi^- \psi(3686)$~\cite{BESIII:2021njb}, $\pi^+ \pi^- \psi_2(3823)$~\cite{BESIII:2022yga}, $\phi \chi_{c1/2}$~\cite{BESIII:2022wjl}, $K^0_SK^0_SJ/\psi$~\cite{Ablikim:2022yav}, $D^{*0} D^{*-} \pi^+$~\cite{BESIII:2023cmv}, and $\Lambda_c^{+}\bar{\Lambda}_c^{-}$~\cite{BESIII:2023rwv}, with the data collected at center-of-mass (c.m.) energies from 4.60 to 4.95 GeV. Except for the $\Lambda_c^{+}\bar{\Lambda}_c^{-}$ channel, these c.m.~energy-dependent cross sections either has signals of or evidence for a structure around the $Y(4660)$ resonance. Evidence of $e^+ e^- \to Y(4660) \to \phi \chi_{c2}$ would indicate a substantial strange quark component in the $Y(4660)$ as suggested by Refs.~\cite{Lebed:2016yvr,Deng:2019dbg,Lu:2016cwr}. But no conclusion can be drawn with the current experimental results and further results on this final state, as well as $\phi \chi_{c0}$, are desirable.   

In this paper, we present a study of the $e^{+}e^{-} \to \phi\chi_{c0}$ process with 6.7 fb$^{-1}$~\cite{lum1,lum2} of data taken at c.m.~energies from 4.47 to 4.95 GeV. The energy-dependent cross section of $e^{+}e^{-} \to \phi\chi_{c0}$  and the process $Y(4660) \to \phi\chi_{c0}$ are studied for the first time. The decays $\phi \to K^{+}K^{-}$, and $\chi_{c0} \to \pi^{+}\pi^{-},\ \pi^{+}\pi^{-}\pi^{0}\pi^{0},\ \pi^{+}\pi^{-}K^{+}K^{-},\  2(\pi^{+}\pi^{-})$, and $3(\pi^{+}\pi^{-})$ are used to reconstruct the signal events.

Using the subset of data samples with $\sqrt{s} \geq 4.84$ GeV, we also search for the long anticipated $J^{PC} = 2^{-+}$ $\eta_{c2}(1D)$ state in $e^{+}e^{-} \to \phi\eta_{c2}(1D)$.  Its predicted mass ranges from 3.80 GeV/$c^2$ to 3.88 GeV/$c^2$ by potential models~\cite{Godfrey:1985xj,Fulcher:1991dm,Zeng:1994vj,Ebert:2002pp,Eichten:2004uh,Barnes:2005pb,Li:2009zu}, or about $3.822$ GeV/$c^2$ based on known masses of other $1^3D_{J}$ states using $M_{\eta_{c2}(1D)} \approx (3M_{\psi(3770)}+5M_{\psi2(1D)}+7M_{X(3842)})/15$~\cite{Belle:2020esr}. With a mass between the $D\bar{D}$ and $D^\ast \bar{D}$ thresholds, $\eta_{c2}(1D)$ lacks open-charm decay channels. Estimated partial decay widths include $\Gamma(\eta_{c2}(1D) \to h_c \gamma) = 303 $ keV, $\Gamma(\eta_{c2}(1D) \to gg) = 110 $ keV, and $\Gamma(\eta_{c2}(1D) \to \pi^+ \pi^- \eta_c) = 45 $ keV~\cite{Eichten:2002qv}. Other predictions suggest a total width of $660-810$ keV, with partial width to light hadrons of $274-392$ keV~\cite{Fan:2009cj}. While the $\eta_{c2}(1D)$ has been sought in $B$ meson decays~\cite{Belle:2020esr} and radiative transitions~\cite{Belle:2021cjk}, it remains the sole unobserved conventional charmonium state without open-charm decays. In the search for $\eta_{c2}(1D)$, the same final states and analysis method as those used for $\chi_{c0}$ are employed.

\section{BESIII detector and Monte Carlo}

The BESIII detector records symmetric $e^+e^-$ collisions provided by the BEPCII storage ring~\cite{Yu:IPAC2016-TUYA01} in the c.m.~energy range from 1.84 to 4.95 GeV with a peak luminosity of $1.1 \times10^{33}\;\text{cm}^{-2}\text{s}^{-1}$ achieved at $\sqrt{s} = 3.773\;\text{GeV}$. The cylindrical core of the BESIII detector covers 93\% of the full solid angle and consists of a helium-based multilayer drift chamber~(MDC), a plastic scintillator time-of-flight system~(TOF), and a CsI(Tl) electromagnetic calorimeter~(EMC), which are all enclosed in a superconducting solenoidal magnet providing a 1.0~T magnetic field. The solenoid is supported by an octagonal flux-return yoke with resistive plate counter muon identification modules interleaved with steel. The charged-particle momentum resolution at $1~{\rm GeV}/c$ is $0.5\%$, and the ${\rm d}E/{\rm d}x$ resolution is $6\%$ for electrons from Bhabha scattering. The EMC measures photon energies with a resolution of $2.5\%$ ($5\%$) at $1$~GeV in the barrel (end cap) region. The time resolution in the TOF barrel region is 68~ps, while that in the end cap region is 110~ps. The end cap TOF system was upgraded in 2015 using multi-gap resistive plate chamber technology, providing a time resolution of 60~ps.  About 87\% of the data used here benefits from this upgrade.  

Simulated data samples produced with a {\sc geant4}-based~\cite{GEANT4:2002zbu} Monte Carlo (MC) package, which includes the geometric description of the BESIII detector and the detector response, are used to determine detection efficiencies, evaluate the initial state radiation (ISR) correction factor, $(1 + \delta)_{\rm ISR}$ and estimate backgrounds. For each of the signal processes $e^+e^- \to \phi\chi_{c0}$ and $\phi\eta_{c2}(1D)$, 50 k signal MC events are generated for each c.m.~energy (see Tables~\ref{Tab:summ} and~\ref{Tab:summ1}) with a phase space (PHSP) model. Subsequently, the $\chi_{c0}$ or $\eta_{c2}(1D)$ decays into $\pi^+\pi^-, \pi^+\pi^-\pi^0\pi^0, K^+K^-\pi^+\pi^-, 2(\pi^+\pi^-), 3(\pi^+\pi^-)$ based on PDG BFs and including known intermediate resonances, and the $\phi$ decays into $K^+K^-$. The simulation includes the beam energy spread and ISR in the $e^+e^-$ annihilation modelled with the generator {\sc kkmc}~\cite{Jadach:1999vf} and {\sc evtgen}~\cite{Ping:2008zz}. A helix-parameter correction for charged tracks is applied during the kinematic fits to improve the consistency between the performance of MC and data. The inclusive MC sample includes the production of open charm processes, the ISR production of vector charmonium(-like) states, and the continuum processes incorporated in {\sc kkmc}. The particle decays are modelled with {\sc evtgen} using branching fractions either taken from the Particle Data Group~\cite{Workman:2022ynf}, when available, or otherwise estimated with {\sc lundcharm}~\cite{Yang:2014vra}. Final state radiation from charged final state particles is incorporated using the {\sc photos} package~\cite{Richter-Was:1992hxq}.

\section{$e^{+}e^{-} \to \phi \chi_{c0}$}
\subsection{Event selection}
In order to improve the selection efficiency, two strategies are applied to reconstruct the signals; full reconstruction of the final states, and  partial reconstruction with a missing $K^{\pm}$, $\pi^{\pm}$, or $\pi^0 $. The number of required particles for different $\chi_{c0}$ decay channels (and $\phi \to K^+K^-$) and the reconstruction methods are listed in Table~\ref{Table:particle}.

\begin{table}[!htbp]
	\setlength{\abovecaptionskip}{0pt}
	\setlength{\belowcaptionskip}{8pt}
	\centering
	\normalsize								
	\setlength{\tabcolsep}{0.01pt}				
	\renewcommand{\arraystretch}{1.2}      	
	\caption{Required numbers of daughter particles for different
          reconstruction methods in the five $\chi_{c0}$ decay channels, 
          with $\phi \to K^+K^-$ always.}
	\label{Table:particle}
	\begin{tabular}{l|ccc|ccc|ccc|ccc}
		\hline\hline
		\multirow{2}{4em}{\centering $\chi_{c0}$ Decay}&\multicolumn{3}{c|}{Full}	&\multicolumn{3}{c|}{Miss $K^\pm$}&\multicolumn{3}{c}{Miss $\pi^\pm$}&\multicolumn{3}{|c}{Miss $\pi^0$}\\ \cline{2-13}
		&$K^\pm$&$\pi^\pm$&$\pi^0$&$K^\pm$&$\pi^\pm$&$\pi^0$&$K^\pm$&$\pi^\pm$&$\pi^0$&$K^\pm$&$\pi^\pm$&$\pi^0$  ~\\ \hline
          $\pi^+\pi^-$           & 2&2&0 & 1&2&0 & 2&1&0 & -&-&- ~\\
          \hline 
          $\pi^+\pi^-\pi^0\pi^0$ & 2&2&2 & 1&2&2 & 2&1&2 & 2&2&1 ~\\
          \hline
          $K^+K^-\pi^+\pi^-$     & 4&2&0 & 3&2&0 & 4&1&0 & -&-&- ~\\
          \hline
          $2(\pi^+\pi^-)$        & 2&4&0 & 1&4&0 & 2&3&0 & -&-&- ~\\
          \hline
	  $3(\pi^+\pi^-)$        & 2&6&0 & 1&6&0 & 2&5&0 & -&-&- ~\\ 
	  \hline\hline
	\end{tabular}
\end{table}
 
The charged tracks detected in the MDC are required to be within an angular range of $|\cos\theta| < 0.93$, where $\theta$ is the polar angle between the charged track and the symmetry axis of the MDC. The distance of the closest approach to the interaction point must be less than 10 cm along the beam direction, and 1 cm in the transverse plane. The particle identification (PID) of the charged tracks are determined by probabilities ${\mathcal P}_{(K,\pi,...)}$, which are calculated by combining the ionization energy loss ($\mathrm{d}E/\mathrm{d}x$) information from the MDC and the time-of-flight information. A track satisfying ${\mathcal P}_K > {\mathcal P}_\pi$ and ${\mathcal P}_K > 0.001$ is identified as a kaon candidate. Similarly, a charged track satisfying ${\mathcal P}_\pi > {\mathcal P}_K$ and ${\mathcal P}_\pi > 0.001$ is identified as a pion candidate. In the $\chi_{c0} \to \pi^{+}\pi^{-}K^{+}K^{-}$ channel, there are four charged kaons in the final state. The kaon pair ($K^{+}K^{-}$) formed from the highest momentum kaons is assumed to come from the $\chi_{c0}$ decay, while the other kaon pair is assumed to come from the $\phi$ decay. This assignment is based on the fact that both of $\phi$ and $\chi_{c0}$ are nearly at rest, so the kaons from the $\phi$ decay are slower. No similar ambiguity exists in other channels.
 
The photon candidates are reconstructed from showers deposited in the EMC. The deposited energy of each shower is required to be larger than 25 MeV in the barrel region, $|\cos\theta| < 0.80$, and larger than 50 MeV in the end cap regions,  $0.86 < |\cos\theta| < 0.92$. To suppress electronic noise and showers unrelated to the event, the difference between the EMC time and the event start time is required to be within [0,700]$\,\mathrm{ns}$. 
The number of photons is required to be $N_\gamma < 8$ for the $\chi_{c0} \to \pi^{+}\pi^{-}\pi^{0}\pi^{0}$ mode, and  $N_\gamma < 4$ for all other $\chi_{c0}$ decay modes. To reconstruct $\pi^{0}$ candidates, we test all possible photon-pair combinations by means of a one-constraint (1C) kinematic fit, in which the invariant mass of the photon pair is constrained to the nominal $\pi^0$ mass. Only $\pi^0$ candidates passing the 1C kinematic fit with a $\chi^{2}_{1C} < 200$ are considered. If more candidates than required for a given mode survive, then those with the lowest $\chi^{2}$ are selected. 

\begin{figure*}[!htb]
	\centering
	\includegraphics[angle=0,width=0.3\textwidth]{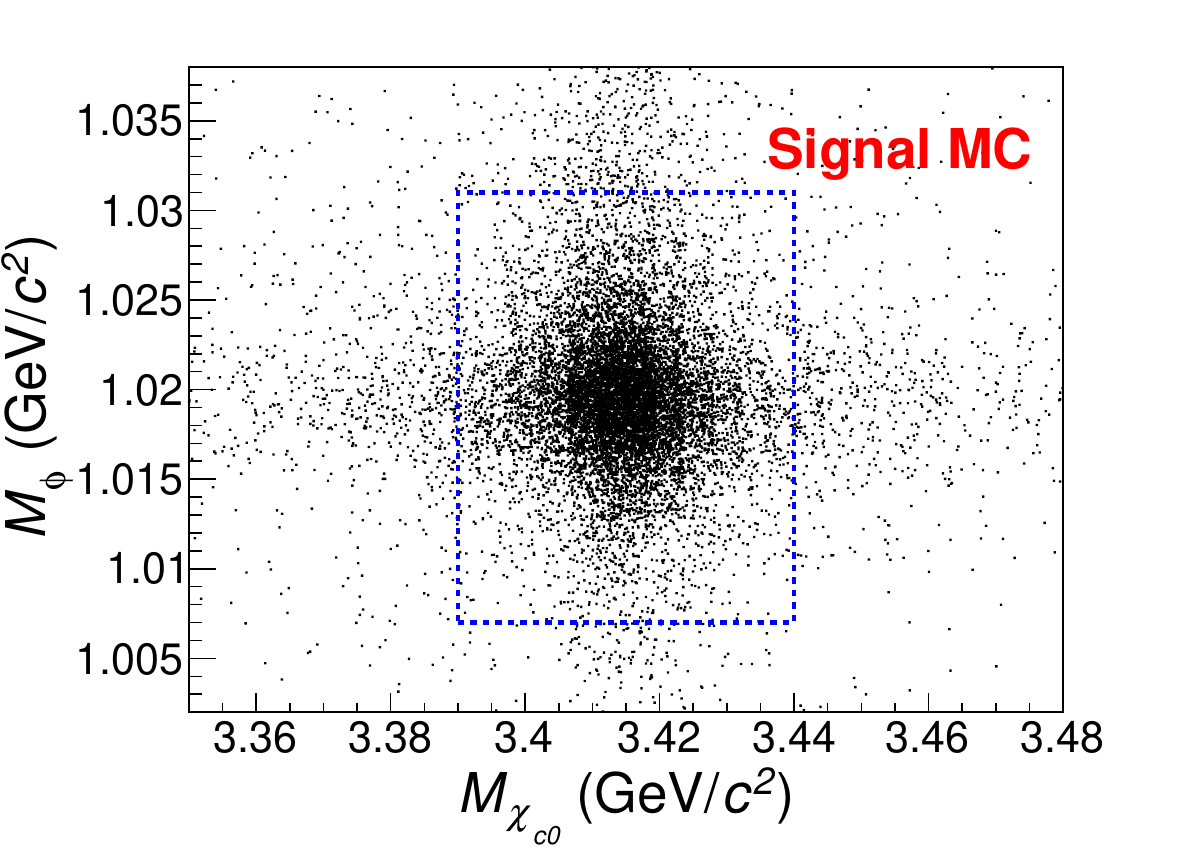}
	\includegraphics[angle=0,width=0.3\textwidth]{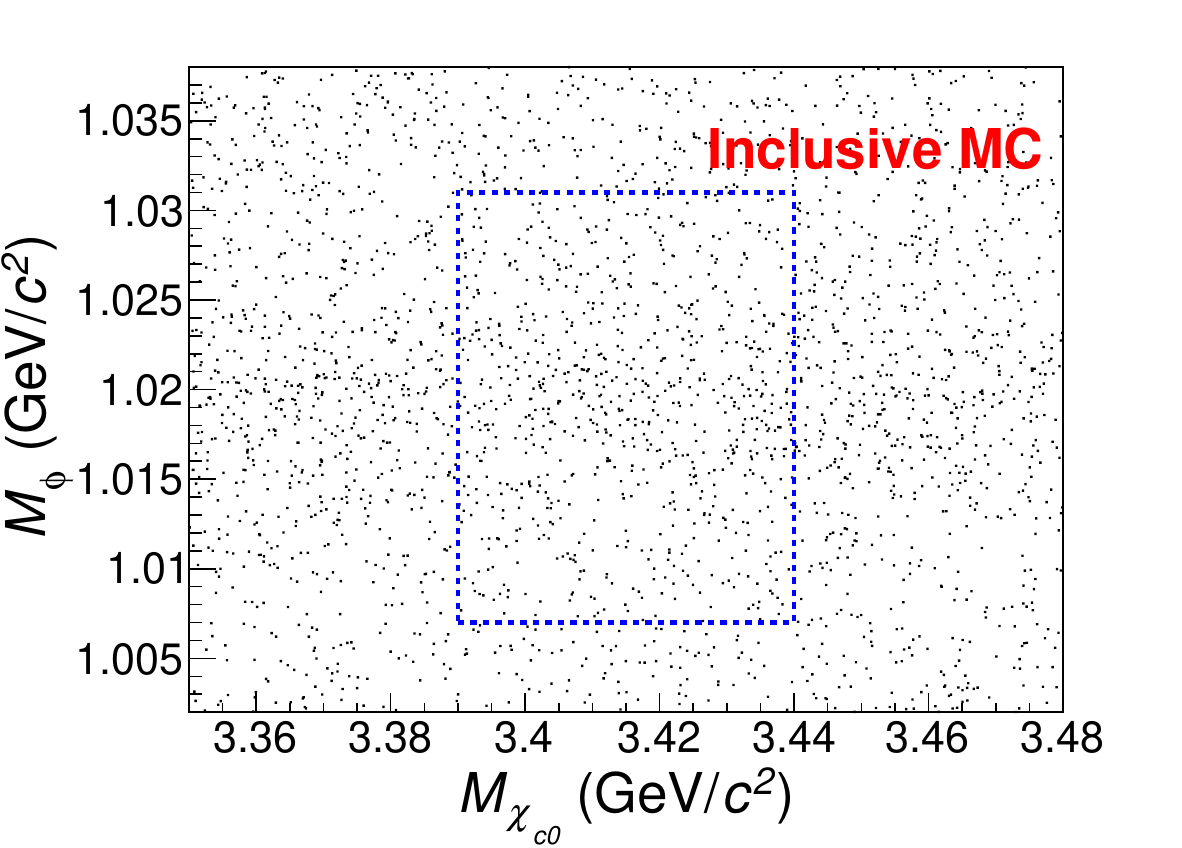}
	\includegraphics[angle=0,width=0.3\textwidth]{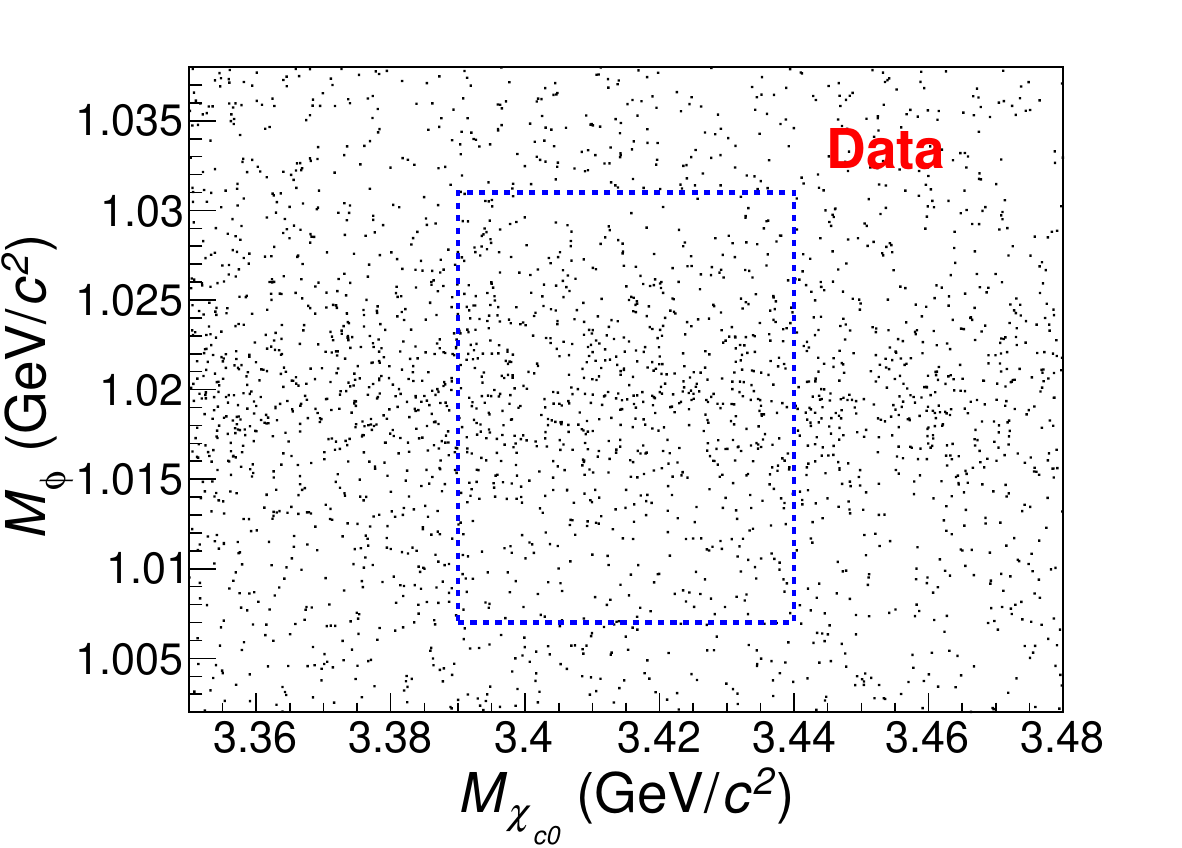}
	\caption{Distributions of $M_{\chi_{c0}}$ versus $M_{\phi}$ for signal MC and inclusive MC at the c.m.~energy of 4.68 GeV, and data, summed over all energy points. Both full and partially reconstructed events are presented.  In this context, $M_{\phi}$ refers to either the invariant mass of $K^+K^-$ or the recoil mass of the $M_{\chi_{c0}}$, and analogously for $M_{\chi_{c0}}$. The dashed boxes indicate the signal regions (though the $\phi$ region is slightly wider for partial reconstruction missing a $K^\pm$ from the $\phi$ decay).}
	\label{Fig:2D}
\end{figure*}

For full-reconstruction modes, the numbers of $\pi$ and $K$ must be exactly as listed in Table~\ref{Table:particle}. A four-constraint (4C) kinematic fit is performed imposing four-momentum conservation on the final state. The signal regions of $\phi$ and $\chi_{c0}$ are set as $1.007 \le M_{\phi} \le$ 1.031 GeV/$c^{2}$ and $3.39 \le M_{\chi_{c0}} \le 3.44$ GeV/$c^{2}$, respectively, as shown in Fig.~\ref{Fig:2D}. Here, the $M_\phi$ indicates the invariant mass of $K^{+}K^{-}$ from the $\phi$ decay and $M_{\chi_{c0}}$ indicates the invariant mass of the candidate $\chi_{c0}$ decay products. The kinematic fit $\chi^{2}$ is required to be less than 40 according to the optimization by the Figure-of-Merit, $FOM = S/({\alpha/2 + \sqrt{B}}$)~\cite{chiopt}, where $\alpha$ represents the expected significance and is set $\alpha = 3$ here. $S$ and $B$ represent the numbers of signal and background events, obtained via the signal and inclusive MC samples, respectively.

For the partial reconstruction mode, the numbers of required particles are listed in Table~\ref{Table:particle}. The net charge is required to be $\mp$1 for the methods with one $K^{\pm}$ or $\pi^{\pm}$ missing, and zero for the method with one $\pi^{0}$ missing. Event selection criteria are different among the three cases.  For modes with a missing $\pi^0$, a 1C kinematic fit is performed by constraining the recoil mass of the $K^+K^-\pi^+\pi^-\pi^0$ system to the $\pi^{0}$ nominal mass. The signal regions of $\phi$ and $\chi_{c0}$ are defined as $1.007 \le M_{\phi} \le 1.031$ GeV/$c^{2}$ and $3.39 \le RM_\phi \le 3.44$ GeV/$c^{2}$, where $RM_{\phi}$ stands for the recoil mass $RM_\phi = \sqrt{p^2_{\rm tot} - p^2_{\phi}}$; here, $p_{\rm tot}$ and $p_{\phi}$ are the four-momenta of the total system and the meson $\phi$, respectively. We require the kinematic fit $\chi^{2} < 6$.   For the modes with a missing $K^\pm$, a 1C kinematic fit is performed by constraining the recoil mass of the $\phi\pi^+\pi^-K^{\pm}$ or $K^{\pm}\chi_{c0}$ to the nominal kaon mass. If the missing kaon is from the $\phi$ decay, the $\phi$ mass window is defined as $1.002 \le RM_{\chi_{c0}} \le 1.038$ GeV/$c^{2}$, otherwise it is $1.007 \le M_{\phi} \le 1.031$ GeV/$c^{2}$. The $\chi_{c0}$ mass window is defined as $3.39 \le M_{\chi_{c0}}$ (or $RM_{\phi}$) $\le 3.44$ GeV/$c^{2}$ in both cases. The kinematic fit must have $\chi^{2}<5$ if the missing kaon is from the $\phi$ decay; otherwise, $\chi^2<45$ is required. For modes with one missing $\pi^\pm$, a 1C kinematic fit is performed by constraining the $RM_{\pi}$ to the charged pion nominal mass. The signal regions of $\phi$ and $\chi_{c0}$ are defined as $1.007 \le M_{\phi} \le 1.031 \mathrm{GeV/c^2}$ and $3.39\le RM_{\phi}\le 3.44 \mathrm{GeV/c^2}$, respectively (as shown in the dashed box in Fig.~\ref{Fig:2D}). The corresponding kinematic fit requirement is $\chi^{2}<30$.  All such $\chi^2$ requirements have been optimized similarly to the full reconstruction mode requirement discussed earlier.  

Figure~\ref{Fig:2D} shows the distributions of $M_{\chi_{c0}}$ versus $M_{\phi}$ for signal MC, inclusive MC, and data over all decay modes and all reconstruction modes. For signal MC and inclusive MC, since the distributions are similar at each energy point, only the sample from $4.68$ GeV is presented as an example; for data, a sum of samples over all energy points is presented to enhance the statistics. The distribution of inclusive MC is similar to that of data. According to the studies of the inclusive MC, the main background events are the continuum processes with multiple kaons and pions in the final state and the open-charm processes. No obvious signal is observed in data.

\subsection{Cross section}
\begin{figure}[!htbp]
	\centering
	\includegraphics[angle=0,width=0.4\textwidth]{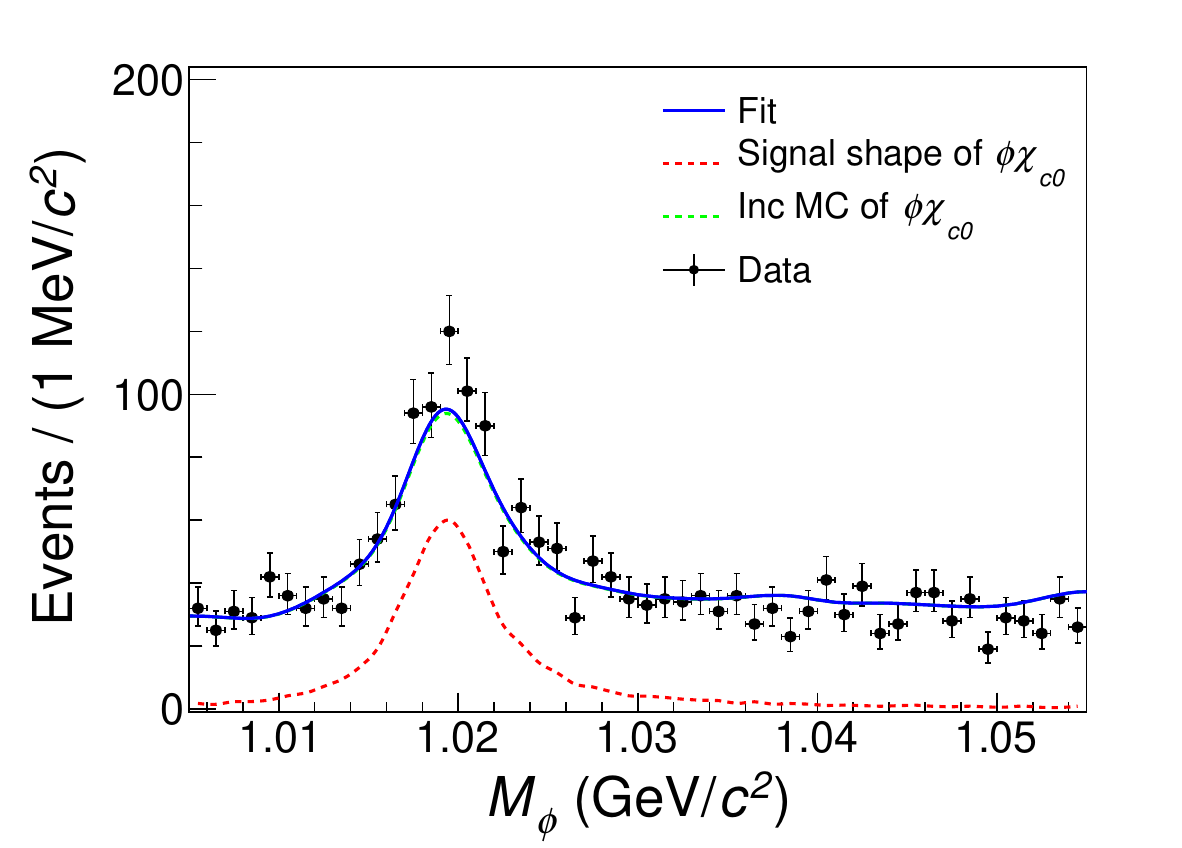}
	\includegraphics[angle=0,width=0.4\textwidth]{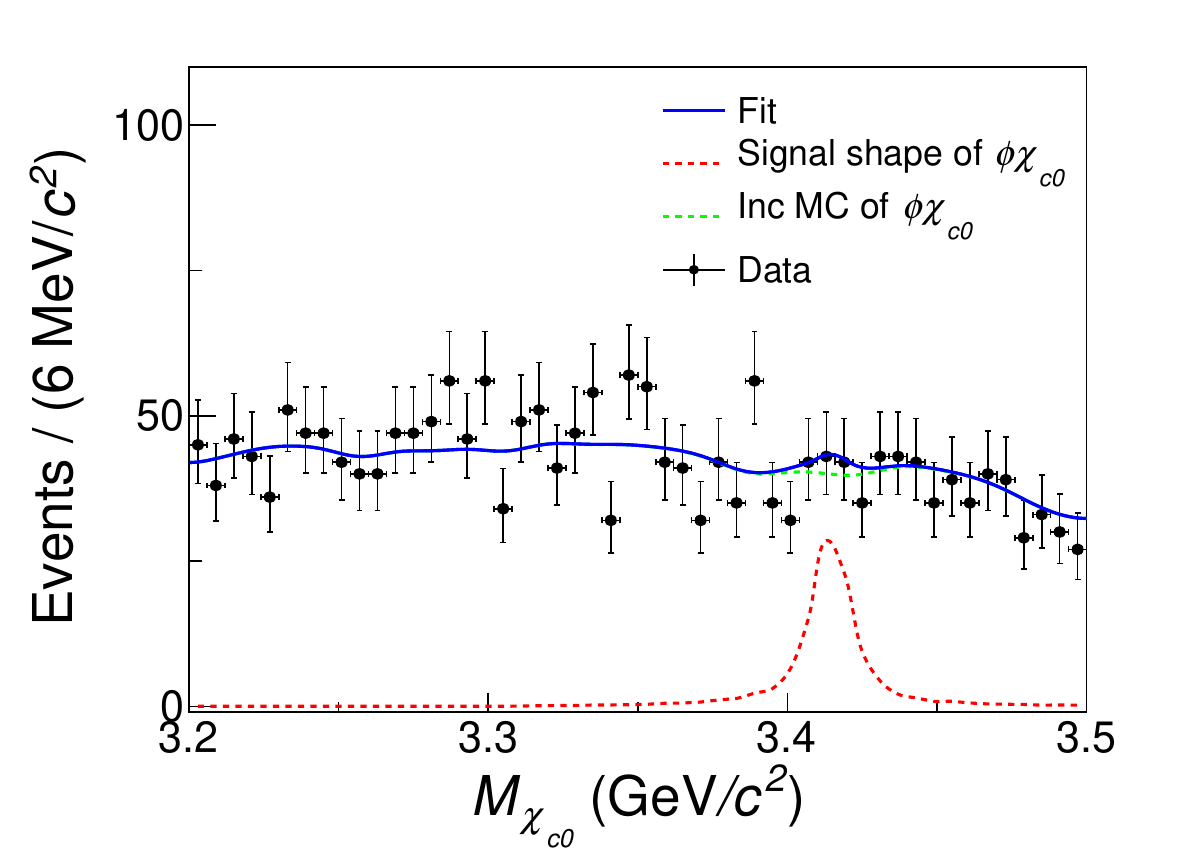}
	\caption{Projections of $M_{\phi}$ (upper) and $M_{\chi_{c0}}$ (lower)
          from the 2D fit at the c.m.~energy of 4.68 GeV.
          The dots with error bars are the data,
          the blue curves are the fit results,
          the green dashed lines are the background components,
          and the red dashed lines represent the signal shapes.
          Since the signal yield is very small, the signal shapes are
          presented with arbitrary normalization.}
	\label{Fig:fitting_result}
\end{figure}

\begin{figure}[!htbp]
	\centering
	\includegraphics[angle=0,width=0.45\textwidth]{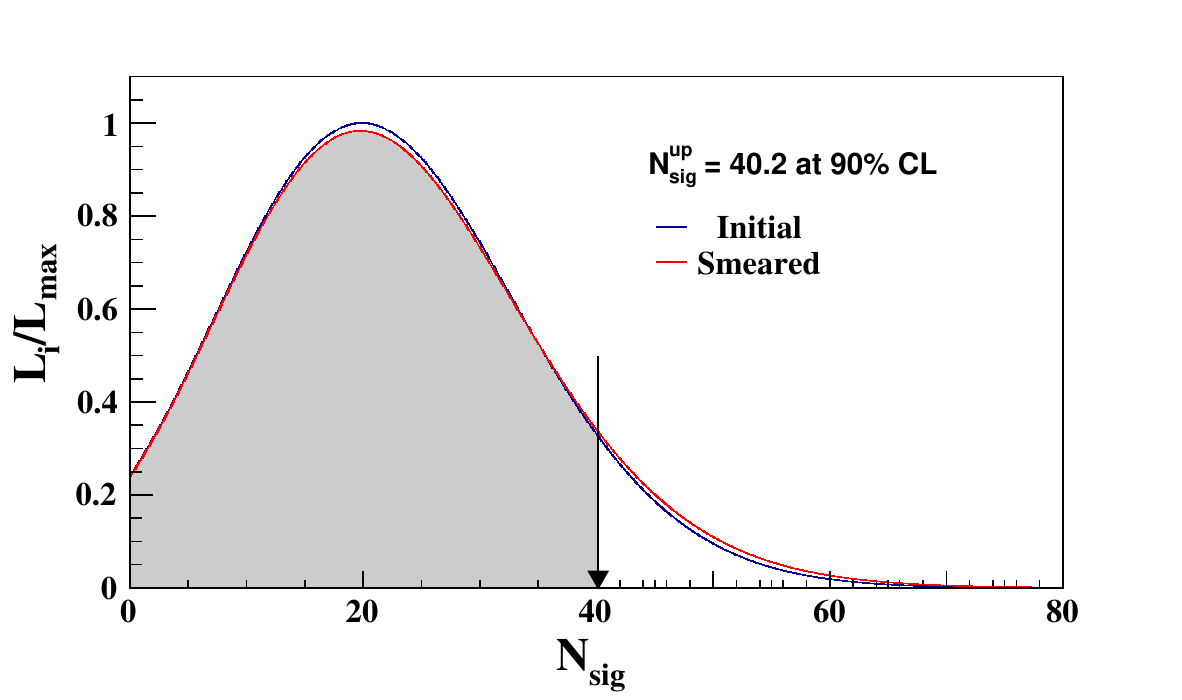}
	\caption{The distribution of normalized likelihood $\rm L_{\rm i}/\rm L_{\rm max}$ versus the signal yield at the c.m.~energy of 4.68 GeV. The blue line is the likelihood profile, the red line is the likelihood with the systematic uncertainty taken into account, and the black arrow indicates the upper limit signal yield at 90\% C.L.}
	\label{Fig:upper_4740}
\end{figure}

The signal yield is extracted by performing a two-dimensional (2D) unbinned maximum-likelihood fit to the distributions of $M_{\phi}$ versus $M_{\chi_{c0}}$. In this fit, the signal shape is derived from the signal MC, and the background shape is derived from the inclusive MC. The statistical significance of the signal is obtained by comparing the log-likelihood values with and without the signal in the fit, after considering the difference in the degrees of freedom. The statistical significance is less than 2$\sigma$ at most c.m.~energies. As an example, Fig.~\ref{Fig:fitting_result} shows the projections of the fit to data for a c.m.~energy of 4.68 GeV. Since no obvious signal is observed, we determine the upper limit of the signal yield at the 90\% confidence level (C.L.) with a Bayesian method~\cite{Stenson:2006gwf} considering the systematic uncertainties discussed in Sec.~\ref{sec:sys}. As an illustration, Fig.~\ref{Fig:upper_4740} shows the distribution of likelihood versus the signal yield at the c.m.~energy of 4.68 GeV, in which the prior function is a step-function at zero to incorporate the fact that physical count rates are always positive.\\

The Born cross section of the process $e^{+}e^{-} \to \phi\chi_{c0}$ at each c.m.~energy $\sqrt{s}$ is determined by
\begin{equation} \label{equ: sigma}
	\sigma^{B}(\sqrt{s}) = \frac{N_{\rm sig}}{\mathcal{L}_{\rm int} \sum \varepsilon^i \, \mathcal{B}_{\chi_{c0}}^i \, \mathcal{B}_{\phi} \, (1+\delta)_{\rm ISR} \, \frac{1}{|1-\Pi|^{2}}  } \
	,
\end{equation} 
where $N_{\rm sig}$ is the signal yield, $\mathcal{L}_{\rm int}$ is the integrated luminosity, $\varepsilon^i$ is the selection efficiency including all reconstruction modes for the $i$th  $\chi_{c0}$ decay channel, $\mathcal{B}_{\chi_{c0}}^i$ is the $i$th branching fraction of $\chi_{c0}$ decay channel (for $\pi^{+}\pi^{-}\pi^{0}\pi^{0}$, there is an additional product term  with the square of the $\mathcal{B}_{\pi^0 \to \gamma \gamma}$) based on the world-average values~\cite{Workman:2022ynf}, $\mathcal{B}_{\phi}$ is the branching fraction of $\phi \to K^{+}K^{-}$~\cite{Workman:2022ynf}, $(1+\delta)_{\rm ISR}$ is the iteratively-determined ISR correction factor, accounting for the line shape of the $e^{+}e^{-} \to \phi \chi_{c0}$ cross sections,  
and $\frac{1}{|1-\Pi|^{2}}$ is the vacuum polarization factor~\cite{WorkingGrouponRadiativeCorrections:2010bjp}. Table~\ref{Tab:summ} lists the obtained cross sections, upper limits, and other quantities used in the calculations for each c.m.~energy.

\begin{table*}[htp]
	\centering
	\setlength{\abovecaptionskip}{0pt}
	\setlength{\belowcaptionskip}{2pt}
	\normalsize 								
	\setlength{\tabcolsep}{2.8pt}				
	\renewcommand{\arraystretch}{1.3}      	
	\caption{\small 
		The measured cross sections and upper limits at each energy point, along with the quantities used in their calculation. Here, $\sqrt{s}$ is the c.m.~energy, $\mathcal{L}_{\rm int}$ is the integrated luminosity, $\bar{\varepsilon}$ is the averaged efficiency $\sum \varepsilon^i \mathcal{B}^i/\sum\mathcal{B}^i$, $N_{\rm sig}$ is the signal yield, $N^{\rm up}_{\rm sig}$ is the upper limit signal yield at 90\% C.L., $N^{\rm up}_{\rm F}$ is the upper limit signal yield at 90\% C.L. after considering multiplicative and additive systematic uncertainty, $(1+\delta)_{\rm ISR}$ is the radiative correction factor, $\frac{1}{|1-\Pi|^{2}}$ is the vacuum polarization factor, $\sigma^{B}$ is the Born cross section, where the first uncertainties are statistical and the second systematic, and $\sigma^{\rm UL}$ is the upper limit including both statistical and systematic uncertainties.}
	\begin{tabular}{c|c|c|c|c|c|c|c|c|c}
		\hline
		\hline
		$\sqrt{s}$~(GeV)	& $\mathcal{L}_{\rm int}$~(pb$^{-1}$) & $\bar{\varepsilon}~(\%)$	 & $N_{\rm sig}$ & $N^{\rm up}_{\rm sig}$ &$N^{\rm up}_{\rm F}$& $(1+\delta)_{\rm ISR}$ & $\frac{1}{|1-\Pi|^{2}}$ &$\sigma^{B}$~(pb)& $\sigma^{\rm UL}$~(pb) ~\\
		\hline
		4.470  &111.1&10.4&$-5.0_{-0.8}^{+1.6}$& 4.1&6.2&0.822&1.055&$-10.6_{-1.7}^{+3.4}\pm1.3$&13.2 ~\\
		4.530  &112.1&16.5&$-4.8_{-1.2}^{+2.1}$ &4.8&6.2&0.874&1.054&$-6.0_{-1.5}^{+2.6}\pm0.6$&7.7 ~\\
		4.575  &48.9&21.1&$0.8_{-1.5}^{+2.4}$ &6.4&8.1&0.779&1.055&$2.0_{-3.8}^{+6.0}\pm0.2$&20.3 ~\\
		4.600  &586.9&22.5&$10.4_{-7.4}^{+8.5}$ &23.8&26.2&0.775&1.055&$2.0_{-1.5}^{+1.7}\pm0.2$&5.2 ~\\
		4.612  &103.8&22.8&$-0.8_{-1.5}^{+2.6}$  &6.2&7.1&0.777&1.055&$-0.9_{-1.6}^{+2.9}\pm0.1$&7.8 ~\\
		4.620  &521.5&23.3&$16.4_{-7.4}^{+8.4}$ &29.4&32.1&0.797&1.055&$3.4_{-1.5}^{+1.7}\pm0.3$&6.7 ~\\
		4.640  &552.4&24.4&$17.3_{-8.2}^{+9.3}$ &31.4&34.0&0.838&1.055&$3.1^{+1.7}_{-1.5}\pm0.3$&6.1 ~\\
		4.660  &529.6&26.1&$26.6_{-8.6}^{+9.5}$ &40.9&40.9&0.825&1.054&$4.7^{+1.7}_{-1.5}\pm0.4$&7.2 ~\\
		4.680  &1669.3&26.0&$11.7_{-11.6}^{+12.8}$ &31.5&40.2&0.936&1.054&$0.6^{+0.6}_{-0.6}\pm0.1$&2.0 ~\\
		4.700  &536.5&26.9&$8.9_{-6.9}^{+8.0}$ &21.7&26.5&0.995&1.055&$1.2^{+1.1}_{-1.0}\pm0.1$&3.7 ~\\
		4.740  &164.3&27.8&$0.6_{-3.0}^{+4.3}$  &9.6&10.4&1.031&1.055&$0.3^{+1.8}_{-1.3}\pm0.0$&4.5  ~\\
		4.750  &367.2&28.3&$-0.9_{-3.4}^{+4.7}$  &9.6&13.4&1.016&1.055&$-0.2^{+0.9}_{-0.6}\pm0.0$&2.6  ~\\
		4.780  &512.8&29.7&$-0.5_{-5.3}^{+6.6}$ &12.8&15.1&0.897&1.055&$-0.1^{+1.0}_{-0.8}\pm0.0$&2.2 ~\\
		4.840  &527.3&30.2&$22.0_{-8.0}^{+9.1}$  &35.8&38.2&0.855&1.056&$3.3^{+1.3}_{-1.2}\pm0.3$&5.7 ~\\
		4.914  &208.1&31.7&$3.8_{-4.5}^{+5.5}$  &13.6&20.3&0.819&1.056&$1.4^{+2.1}_{-1.7}\pm0.1$&7.6 ~\\
		4.946  &160.3&31.6&$-4.8_{-1.8}^{+2.8}$  &5.6&7.0&0.818&1.056&$-2.3^{+1.4}_{-0.9}\pm0.2$&3.4 ~\\
		\hline \hline
	\end{tabular}
	\label{Tab:summ}
\end{table*}

\subsection{Systematic uncertainty}
\label{sec:sys}
In the upper limit determination of the cross sections, the systematic uncertainties for $e^{+}e^{-} \to \phi\chi_{c0}$ are classified into multiplicative and additive categories. Multiplicative uncertainties are associated with tracking and photon reconstruction, integrated luminosity, PID, the kinematic fit, external branching fractions, and the ISR correction. Additive uncertainties are associated with the 2D fit procedure.

The uncertainty associated with the tracking and PID for each charged track is taken as 1.0\%~\cite{BESIII:2013qmu}, and the uncertainty due to photon reconstruction for each photon is taken as 1.0\%~\cite{BESIII:2010ank}. The total systematic uncertainty from the tracking is estimated by lowering of 1\% the tracking efficiency for each track in the reconstruction process for the MC samples. The difference in the efficiencies obtained with and without the 1\% lowering is taken as the systematic uncertainty associated with the charged track reconstruction. The systematic uncertainty due to the photon reconstruction is obtained by the same method. The uncertainty associated with PID is obtained by switching the hypothesis of each charged kaon and pion randomly with 1.0\% probability~\cite{BESIII:2013qmu}, the difference between the efficiencies with and without the switching is taken as the uncertainty associated with the PID. The integrated luminosity is measured using large-angle Bhabha events with an uncertainty of 1.0\%~\cite{lum1}. The efficiency after the helix correction is adopted as the nominal one, and half of the difference from the uncorrected value is adopted as the corresponding uncertainty.  The uncertainties of the branching fractions of $\phi \to K^{+}K^{-}$, $\chi_{c0}\to\pi^{+}\pi^{-}$, $\pi^{+}\pi^{-}\pi^{0}\pi^{0}$, $\pi^{+}\pi^{-}K^{+}K^{-}$, $2(\pi^{+}\pi^{-})$ and $3(\pi^{+}\pi^{-})$ are taken from the PDG~\cite{Workman:2022ynf}. The uncertainty of the averaged efficiency caused by these uncertain branching fractions is determined by calculating the averaged efficiency 1000 times, with the branching fractions following a Gaussian distribution corresponding to the branching fraction knowledge. The efficiency distribution is fitted with a Gaussian function and the width is taken as the systematic uncertainty. The uncertainty due to the ISR correction factor is estimated by varying the nominal PHSP line shape to one with an added $Y(4660)$ signal. The differences in the ISR correction factors are taken as the systematic uncertainties. Table~\ref{Tab:sys_total} summarizes all the multiplicative systematic uncertainties; the total is obtained by summing the individual items in quadrature. Note that these uncertainties are energy-dependent, but the variation is very small. Therefore, averaged values are adopted. 

\begin{table}[!htbp]
	\setlength{\abovecaptionskip}{0pt}
	\setlength{\belowcaptionskip}{8pt}
	\centering
	\normalsize 								
	\setlength{\tabcolsep}{17pt}				
	\renewcommand{\arraystretch}{1.2}      	
	\caption{The multiplicative systematic uncertainties, in \%, for the cross section measurements of $e^{+}e^{-} \to \phi\chi_{c0}$. Here, $\mathcal{B}_{\rm eff}$ is the uncertainty of the averaged efficiency caused by the uncertainty associated with the branching fractions.}
	\begin{tabular}{c | c}
		\hline\hline
		Source						&   Uncertainty  	~\\
		\hline
		Tracking  					& 	2.7 ~\\
		Photon reconstruction	&	0.3	~\\
		PID							&	4.6	~\\
		Luminosity					&   1.0 ~\\
		$\mathcal{B}_{\rm eff}$		&	0.9 ~\\	
		$\mathcal{B}_{\chi_{c0}}$	&	5.3	~\\
		$\mathcal{B}_\phi$			&	1.0 ~\\
		Kinematic fit				&	0.1 ~\\
		ISR correction		&	5.5  ~\\
		\hline
		Total						&	9.5  ~\\
		\hline \hline
	\end{tabular}
	\label{Tab:sys_total}
\end{table}

The additive systematic uncertainties of the 2D fit are associated with signal shape,  background shape, and fitting range. The systematic uncertainty caused by the signal shape, which is mainly from the resolution difference between data and MC simulation, is negligible based on the study of control samples. In the nominal fit, the background shape is extracted from the inclusive MC, which has two kinds of backgrounds, with or without a $\phi$ signal. We change to a new shape using a 2D 2nd-order polynomial function and a shape extracted from a mix of MC samples $e^{+}e^{-} \to \phi X$, where $X$ includes final states $\pi^{+}\pi^{-}$, $\pi^{+}\pi^{-}\pi^{0}\pi^{0}$, $\pi^{+}\pi^{-}K^{+}K^{-}$, $K^{+}K^{-}K^{+}K^{-}$, $2(\pi^{+}\pi^{-})$, and $3(\pi^{+}\pi^{-})$ without any intermediate resonances. The proportions of these six final states in the mixing MC sample are fixed to the branching fractions of $\chi_{c0}$ decay channels from the PDG~\cite{Workman:2022ynf}. The difference between the nominal and new results is taken as the corresponding uncertainty. The systematic uncertainty due to the fit range is examined by varying the fitting ranges of both the $\phi$ and $\chi_{c0}$ by a few MeV. The maximum upper limit among all variations is adopted as the effect of additive systematic uncertainty. Typically, the upper limits of the signal yields increase in $(10-20)\%$ after considering the additive systematic uncertainty.

\section{Search for $e^{+}e^{-} \to Y(4660) \to \phi\chi_{c0}$}
\begin{figure}[!htbp]
	\centering
	\includegraphics[angle=0,width=0.4\textwidth]{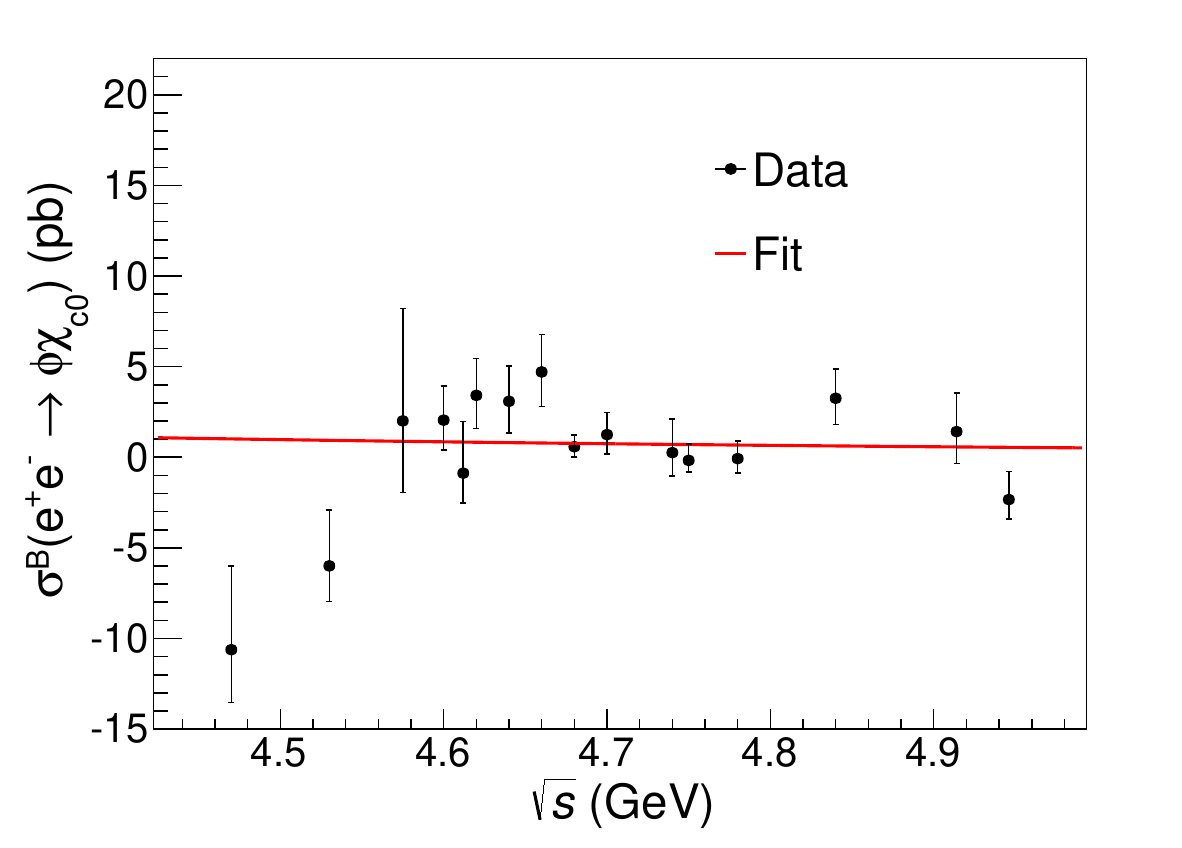}
	\includegraphics[angle=0,width=0.4\textwidth]{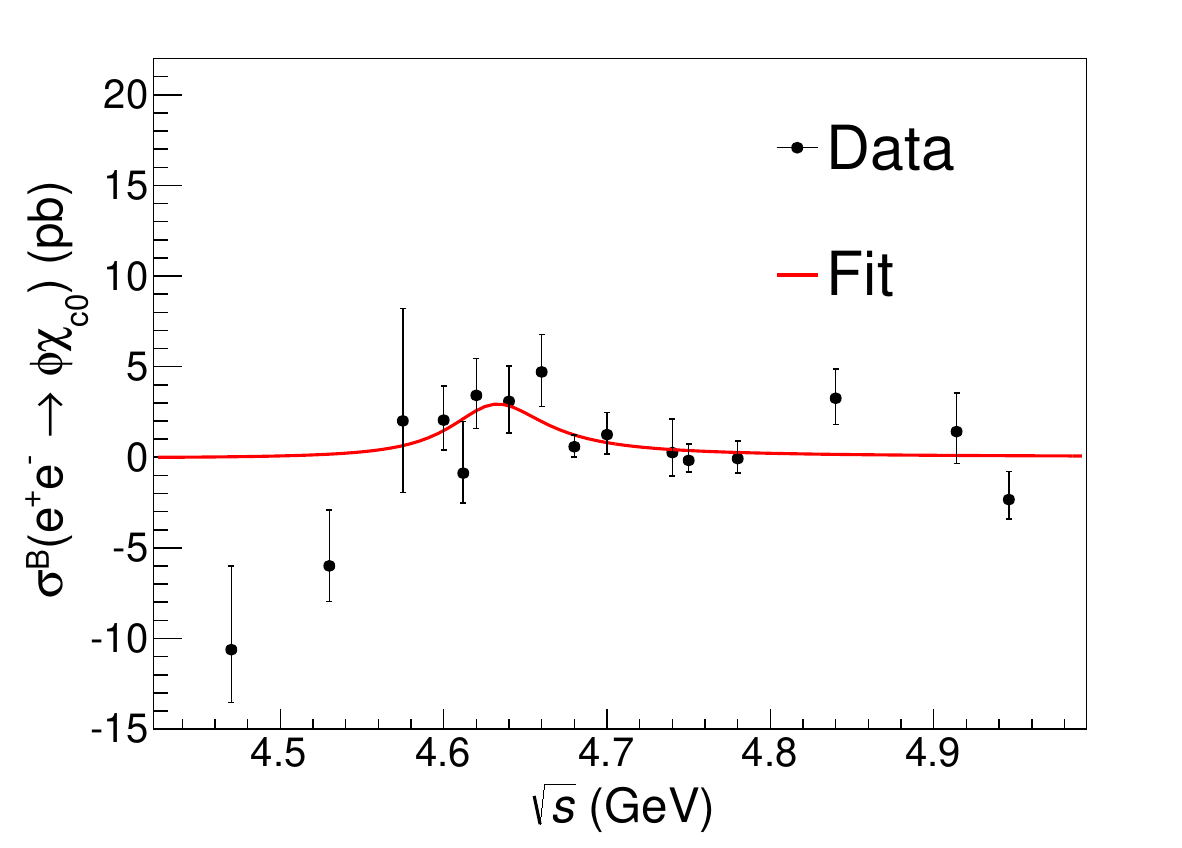}
	\caption{Fit to the Born cross sections, with combined statistical and systematic uncertainties, for $e^{+}e^{-} \to \phi\chi_{c0}$ with a continuum amplitude (top) and a Breit-Wigner function for the $Y(4660)$ (bottom).}
	\label{Fig:fitline}
\end{figure}

Figure~\ref{Fig:fitline} shows the Born cross sections of the process $e^+ e^- \to \phi \chi_{c0}$ with both statistical and systematic uncertainties. In order to study the contribution of $Y(4660)$,  a maximum likelihood fit is performed and an incoherent sum of continuum process and $Y(4660)$ components is utilized to describe the line shape of the Born cross section:
\begin{equation} \label{fitsigma}
	\sigma^{B}(\sqrt{s}) = \frac{1}{|1-\Pi|^{2}} \left( \frac{f}{(\sqrt{s}/M)^{n}} \frac{\Phi(\sqrt{s})}{\Phi(M)} + \left|BW(\sqrt{s})\right|^{2} \right),
\end{equation}
where the Breit-Wigner (BW) function describing the resonance is
\begin{equation}
	BW(\sqrt{s}) = \frac{M}{\sqrt{s}} \frac{\sqrt{12\pi\Gamma_{\rm tot}\Gamma_{e^{+}e^{-}}\mathcal{B}_{\phi\chi_{c0}}}}{s-M^{2} + iM\Gamma_{\rm tot}}  \sqrt{\frac{\Phi(\sqrt{s})}{\Phi(M)}}\ .
\end{equation}
Here, $f$ and $n$ are the parameters for the continuum process; $M$, $ \Gamma_{\rm tot}$, and $\Gamma_{e^{+}e^{-}}$ are the mass, the total width, and the electronic width of the $Y(4660)$, respectively; $\mathcal{B}_{\phi\chi_{c0}}$ is the branching fraction of  $Y(4660) \to \phi\chi_{c0}$ and $\Phi$ is the standard two-body decay ($Y \to \phi\chi_{c0}$) PHSP factor. In this fit,  the significance of the contributions of $Y(4660)$ and continuum are determined to be 2.1$\sigma$ and  0.3$\sigma$, respectively., and this fit is not considered any further.  Instead, we use either or a continuum shape only, or a BW function only, to fit the cross sections. Figure~\ref{Fig:fitline} shows these fit results.  The continuum fit has a poor $\chi^2$ (see Table~\ref{Tab:summarypar}). 
In the BW fit, the mass and total width of $Y(4660)$ are fixed to 4630 MeV and 72 MeV~\cite{Workman:2022ynf}, and the relevant parameters are listed in Table~\ref{Tab:summarypar}. The $\Gamma_{e^{+}e^{-}}\mathcal{B}_{\phi\chi_{c0}}$ is determined to be $0.29 \pm 0.08$ eV. The significance is determined to be 2.4$\sigma$ by comparing the likelihood values in different hypotheses, with or without the BW term, as well as the difference in the number of degrees of freedom. 

\begin{table}[!htbp]
	\centering
	\setlength{\abovecaptionskip}{0pt}
	\setlength{\belowcaptionskip}{8pt}
	\normalsize 								
	\setlength{\tabcolsep}{10pt}				
	\renewcommand{\arraystretch}{1.2}      	
	\caption{Parameters from the two fits to the Born cross sections. The uncertainties are statistical only.}
	\begin{tabular}{c|c|c}
		\hline\hline
		Parameters			&	Continuum &  BW 	~\\
		\hline
		$\Gamma_{e^{+}e^{-}}\mathcal{B}_{\phi\chi_{c0}}$ (eV) &  - & $0.29\pm0.08$      ~\\ \hline
		M(MeV) &4630 (fixed)	 & 4630 (fixed)	 ~\\  \hline
		$\Gamma_{\rm tot}$ (MeV) &-	&72	(fixed) ~\\ \hline
		f &	 $0.3 \pm 0.1$ &-		~\\ \hline
		n &	 $6 \pm 15$ &	-	~\\ \hline
		$\chi^2$/ndf &	29.2/14 &22.2/15	 ~\\
		\hline \hline
	\end{tabular}
	\label{Tab:summarypar}
\end{table}

The upper limit of $\Gamma_{e^{+}e^{-}}\mathcal{B}_{\phi\chi_{c0}}$ is obtained by varying its value and scanning the likelihood distribution of $\Gamma_{e^{+}e^{-}}\mathcal{B}$. We take the value corresponding to the 90\% C.L. as the upper limit of $\Gamma_{e^{+}e^{-}}\mathcal{B}_{\phi\chi_{c0}}$.
The systematic uncertainty from the fitting is determined by varying the value of parameters $M$ and $\Gamma_{\rm tot}$ within the one standard deviation for the mass, i.e., of $(4630 \pm 6)$ MeV, and for the total width, $(72_{-12}^{+14})$ MeV. The maximum difference with respect to the standard value is found to be $0.01$ eV. The product $\Gamma_{e^{+}e^{-}}\mathcal{B}_{\phi\chi_{c0}}$ is determined to be $0.29 \pm 0.08$ eV, and the upper limit at 90\% C.L. is $0.40$ eV.

\section{$e^{+}e^{-} \to \phi \eta_{c2}(1D)$}
While the analysis strategies for the process $e^{+}e^{-} \to \phi \eta_{c2}(1D)$ closely resemble those for $e^{+}e^{-} \to \phi\chi_{c0}$, differences exist in event selection, upper limit determination, and systematic uncertainty estimation.

In the event selection criteria, the mass window for $\eta_{c2}(1D)$ is defined as $3.76 \le M_{\eta_{c2}(1D)} \le 3.88$ GeV/$c^{2}$ to replace the $\chi_{c0}$ mass window. In the full reconstruction mode, the requirement for 4C kinematic fit is optimized to be less than $\chi^2<60$. For partial reconstruction modes, this $\chi^{2}$ requirement optimized to be $<10$, $<20$, $<5$, and $<80$ for a missing $\pi^0$, a missing $\pi^\pm$, a missing $K^\pm$ from the $\phi$, and a missing $K^\pm$ from the $\eta_{c2}(1D)$, respectively. 

\begin{figure}[!htbp]
	\centering
	\includegraphics[angle=0,width=0.4\textwidth]{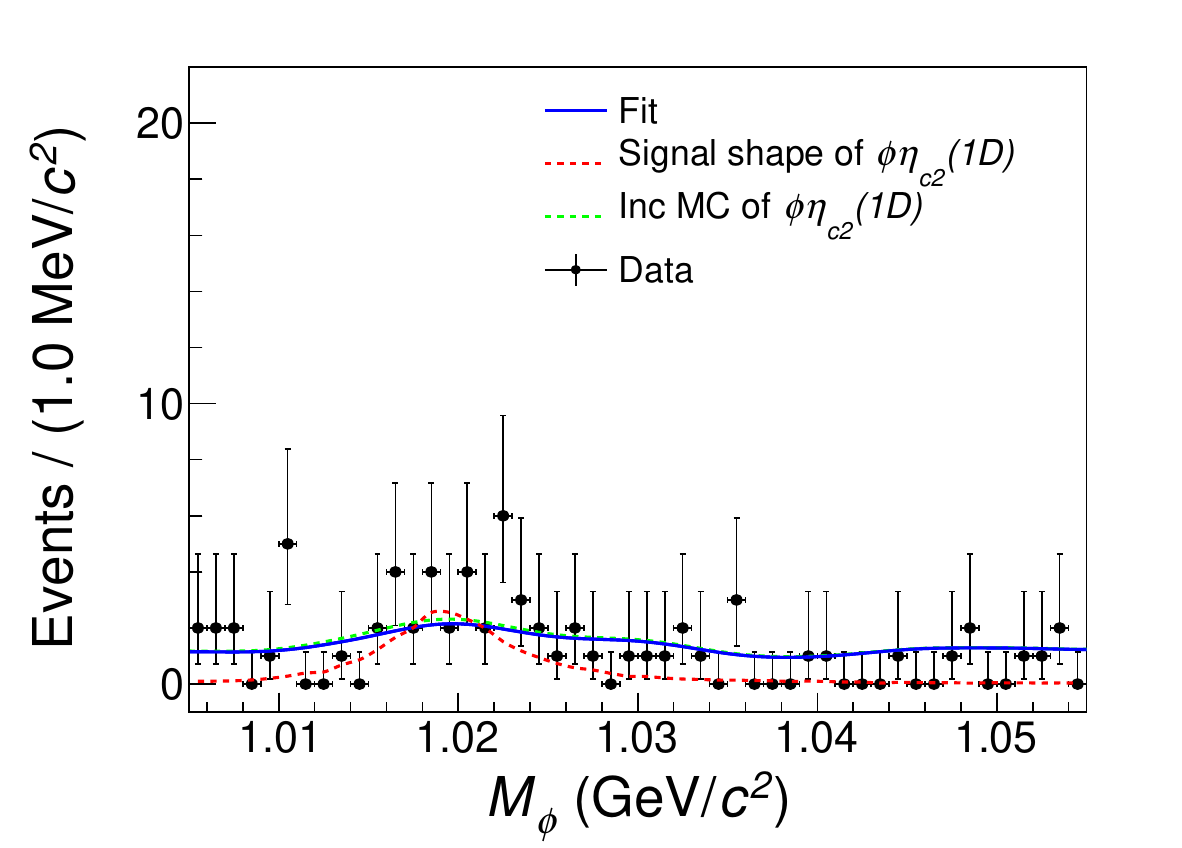}
	\includegraphics[angle=0,width=0.4\textwidth]{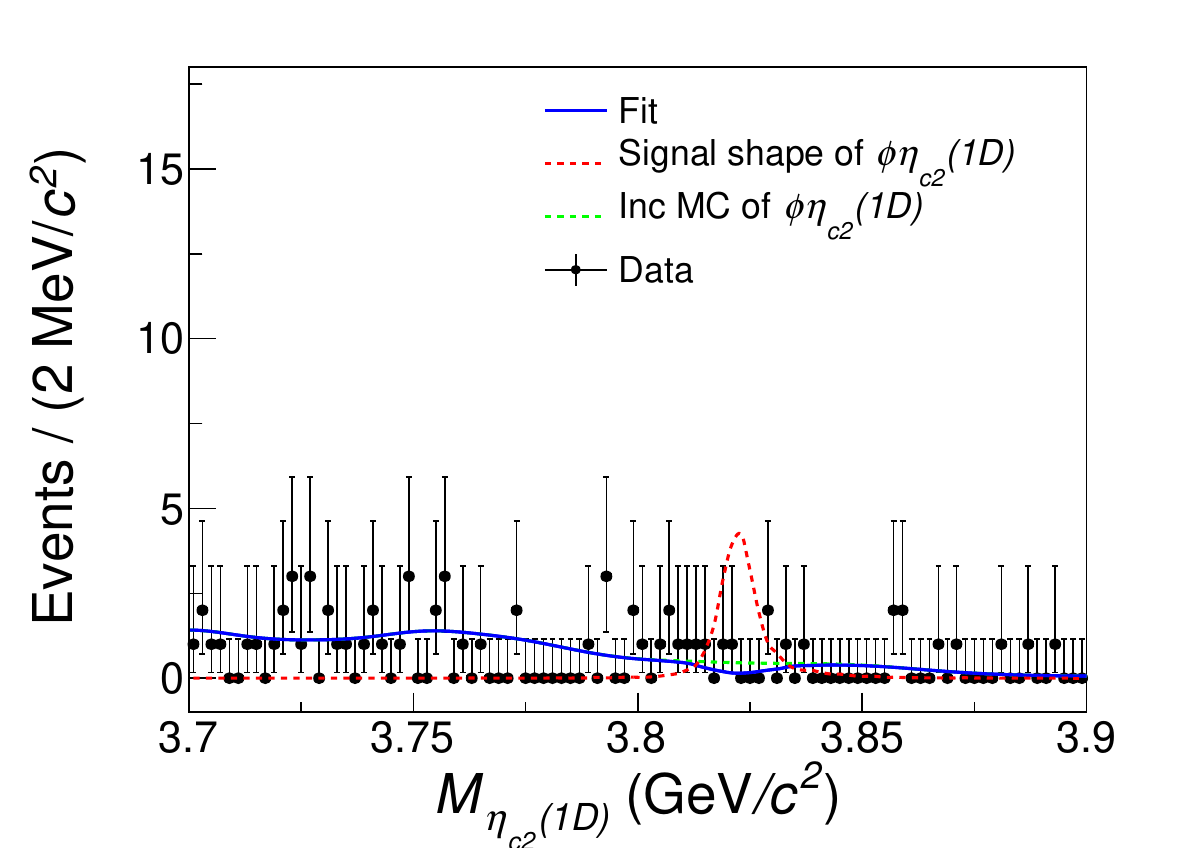}
	\caption{Projections of $M_{\phi}$ (upper) and $M_{\eta_{c2}(1D)}$ (lower) from the 2D fit at the c.m.~energy of 4.914 GeV. The dots with error bars are the data, the blue curves are the fit results, the green dashed lines are the background components, and the red dashed lines represent the signal shapes. Since the signal yield is tiny, the signal shapes are presented with arbitrary normalization.}
	\label{Fig:fitting_result_eta}
\end{figure}

\begin{figure}[!htbp]
	\centering
	\includegraphics[angle=0,width=0.45\textwidth]{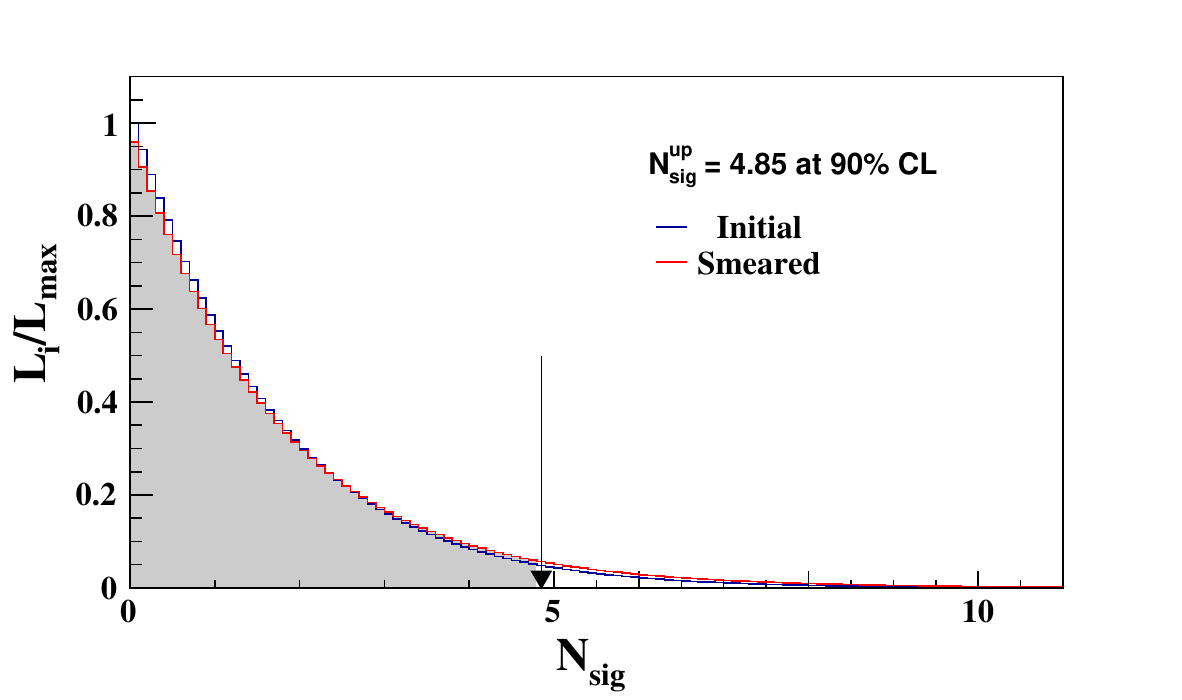}
	\caption{The distribution of normalized likelihood $\rm L_{\rm i}/\rm L_{\rm max}$ versus the signal yield at the c.m.~energy of 4.914 GeV. The blue line is the likelihood profile, the red line is the likelihood with the systematic uncertainty taken into account, and the black arrow indicates the upper limit signal yield at 90\% C.L.}
	\label{Fig:upper1_4914}
\end{figure}

Following the application of all selection criteria, no discernible signal is observed in the data at each energy point. Fit results at the c.m.~energy of 4.914 GeV are depicted in Fig.~\ref{Fig:fitting_result_eta} as an illustration, with the determination of the upper limit signal yield shown in Fig.~\ref{Fig:upper1_4914} after accounting for multiplicative systematic uncertainties.


Given the absence of known branching fractions for the decays of $\eta_{c2}(1D)$, the upper limit on the product of the cross section for the process $e^{+}e^{-} \to \phi\eta_{c2}(1D)$ and the sum of the five branching fractions is determined by
\begin{equation} \label{equ: sigma1}
	\sigma^{\rm U.L.}\mathcal{B}_{\eta_{c2}(1D)} = \frac{N^{\rm U.L.}}{\mathcal{L}_{\rm int} ,\ \bar{\varepsilon} \,  (1+\delta)_{\rm ISR} \, \frac{1}{(1-\Pi)^{2}} \, \mathcal{B}_{\phi}} \
	,
\end{equation}
where several symbols hold the same meaning as in Eq.~(\ref{equ: sigma}), and $\mathcal{B}_{\eta_{c2}(1D)}$ represents the sum of the five $\eta_{c2}(1D)$ branching fractions, and $\bar{\varepsilon}$ denotes the average selection efficiency under the assumption of uniform branching fractions. The results obtained and the quantities utilized in the calculations at each center-of-mass energy are detailed in Table~\ref{Tab:summ1}.

\begin{table*}[!htp]
	\centering
	\setlength{\abovecaptionskip}{0pt}
	\setlength{\belowcaptionskip}{2pt}
	\normalsize 								
	\setlength{\tabcolsep}{2.8pt}				
	\renewcommand{\arraystretch}{1.3}      	
	\caption{\small 
		The upper limits on the product of cross section of the process $e^{+}e^{-} \to \phi\eta_{c2}(1D)$ and the sum of branching fractions, and the quantities used in the calculation at each energy point. Here $\sqrt{s}$ is the c.m.~energy, $\mathcal{L}_{\rm int}$ is the integrated luminosity, $\bar{\varepsilon}$ is the averaged efficiency, $N_{\rm sig}$ is the signal yield, $N^{\rm up}_{\rm sig}$ is the upper limit of signal yield at 90\% C.L., $N^{\rm up}_{\rm F}$ is the upper limit of signal yield at 90\% C.L. after considering the multiplicative and additive systematic uncertainty, $(1+\delta)_{\rm ISR}$ is the radiative correction factor, $\frac{1}{|1-\Pi|^{2}}$ is the vacuum polarization factor, and $\sigma^{\rm U.L.}\mathcal{B}_{\eta_{c2}(1D)}$ is the U.L.~on the product of cross section of the process $e^{+}e^{-} \to \phi\eta_{c2}(1D)$ and a sum of the branching fractions of five decay channels of $\eta_{c2}(1D)$.}
	\begin{tabular}{c|c|c|c|c|c|c|c|c}
		\hline
		\hline
		$\sqrt{s}$~(GeV)	& $\mathcal{L}_{\rm int}$~(pb$^{-1}$) & $\bar{\varepsilon}~(\%)$	 & $N_{\rm sig}$ & $N^{\rm up}_{\rm sig}$ &$N^{\rm up}_{\rm F}$& $(1+\delta)_{\rm ISR}$ & $\frac{1}{|1-\Pi|^{2}}$ & $\sigma^{\rm U.L.}\mathcal{B}_{\eta_{c2}(1D)}$~(pb) ~\\
		\hline
		4.840  &527.3&4.2&$-7.8_{-1.3}^{+1.6}$  &3.6&15.5&0.598&1.056&1.1 ~\\
		4.914  &208.1&14.0&$-6.3_{-1.6}^{+2.2}$  &4.5&11.3&0.765&1.056&0.5 ~\\
		4.946  &160.3&17.2&$-9.6_{-1.5}^{+2.6}$  &4.6&16.6&0.793&1.056&0.7 ~\\
		\hline \hline
	\end{tabular}
	\label{Tab:summ1}
\end{table*}

The primary source of multiplicative uncertainty stems from the lack of knowledge regarding the branching fractions of $\eta_{c2}(1D)$ decays. This uncertainty is estimated by determining the averaged efficiency, under the assumption that the branching fractions of the five decay channels of $\eta_{c2}(1D)$ are equal to the averaged values of those of $\chi_{c0}$ and $\chi_{c2}$. The difference between the newly obtained efficiency and the nominal efficiency is taken as the corresponding uncertainty. The uncertainty arising from the ISR correction factor is evaluated by varying a flat line shape to a PHSP line shape. Table~\ref{Tab:sys_total1} provides a summary of all the multiplicative systematic uncertainties.

\begin{table}[!htbp]
	\setlength{\abovecaptionskip}{0pt}
	\setlength{\belowcaptionskip}{8pt}
	\centering
	\normalsize 								
	\setlength{\tabcolsep}{17pt}				
	\renewcommand{\arraystretch}{1.2}      	
	\caption{The multiplicative systematic uncertainties, in \%, for the cross section measurements of $e^{+}e^{-} \to \phi\eta_{c2}(1D)$.}
	\begin{tabular}{c | c}
		\hline\hline
		Source						&   Uncertainty  	~\\
		\hline
		Tracking  					& 	2.6 ~\\
		Photon reconstruction	&	0.9	~\\
		PID							&	3.6	~\\
		Luminosity					&   1.0 ~\\
		$\mathcal{B}_{\rm eff}$		&	22.3 ~\\	
		$\mathcal{B}_\phi$			&	1.0 ~\\
		Kinematic fit				&	1.4 ~\\
		ISR correction		&	3.5  ~\\
		\hline
		Total						&	23.1  ~\\
		\hline \hline
	\end{tabular}
	\label{Tab:sys_total1}
\end{table}

The additive systematic uncertainties in the 2D fit are linked to the background shape, fitting range, and the mass of $\eta_{c2}(1D)$. The uncertainty pertaining to the width of $\eta_{c2}(1D)$ is disregarded due to the anticipated narrowness of the width. Following a methodology akin to that employed in the $\chi_{c0}$ investigation, the maximum upper limit derived from all possible combinations of variations is chosen as the additive systematic uncertainty. The mass of the $\eta_{c2}(1D)$ is varied from $3.80 \sim3.88$ GeV/$c^2$. 

\section{Conclusion AND DISCUSSION}
In summary, searches for the $e^{+}e^{-} \to \phi\chi_{c0}$ and $e^{+}e^{-} \to \phi\eta_{c2}(1D)$ processes are conducted at center-of-mass energies ranging from 4.47 to 4.95 GeV. No significant signals of either process are observed at any of the center-of-mass energies. The upper limits of the Born cross section at a 90\% confidence level for $e^{+}e^{-} \to \phi\chi_{c0}$ are determined, as well as the product of the cross section of the process $e^{+}e^{-} \to \phi\eta_{c2}(1D)$ and the sum of branching fractions of five channels. At a center-of-mass energy of 4.60 GeV, the upper limit of the cross section for $e^{+}e^{-} \to \phi\chi_{c0}$ is slightly more stringent than the previously reported value~\cite{BESIII:2017qtm}, i.e., $5.2$ pb compared to $5.4$ pb, due to the inclusion of a larger number of decay channels. Additionally, the significance of the signal of the decay $Y(4660) \to \phi\chi_{c0}$ is determined to be 2.4$\sigma$, and $\Gamma_{e^{+}e^{-}}\mathcal{B}_{Y(4660) \to \phi\chi_{c0}}$ and the corresponding upper limit at a 90\% confidence level are determined to be $0.29 \pm 0.08$ eV and 0.40 eV, respectively. These results are notably smaller than those of the $\chi_{c2}$ channel, which was measured to be $\Gamma_{e^{+}e^{-}}\mathcal{B}_{Y(4660) \to \phi\chi_{c2}} = 0.74 \pm 0.13$ eV~\cite{BESIII:2022wjl}. The measured results of this analysis provide valuable information for a better understanding of the vector charmonium-like state $Y(4660)$.

\section*{Acknowledgement}

The BESIII Collaboration thanks the staff of BEPCII and the IHEP computing center for their strong support. This work is supported in part by National Key R\&D Program of China under Contracts Nos. 2020YFA0406300, 2020YFA0406400, 2023YFA1606000; National Natural Science Foundation of China (NSFC) under Contracts Nos. 12035009, 11875170, 11635010, 11735014, 11935015, 11935016, 11935018, 12025502, 12035013, 12061131003, 12192260, 12192261, 12192262, 12192263, 12192264, 12192265, 12221005, 12225509, 12235017, 12361141819; the Chinese Academy of Sciences (CAS) Large-Scale Scientific Facility Program; the CAS Center for Excellence in Particle Physics (CCEPP); Joint Large-Scale Scientific Facility Funds of the NSFC and CAS under Contract No. U1832207; CAS under Contract No. YSBR-101; 100 Talents Program of CAS; The Institute of Nuclear and Particle Physics (INPAC) and Shanghai Key Laboratory for Particle Physics and Cosmology; Agencia Nacional de Investigacin y Desarrollo de Chile (ANID), Chile under Contract No. ANID PIA/APOYO AFB230003; German Research Foundation DFG under Contracts Nos. 455635585, FOR5327, GRK 2149; Istituto Nazionale di Fisica Nucleare, Italy; Knut and Alice Wallenberg Foundation under Contracts Nos. 2021.0174, 2021.0299; Ministry of Development of Turkey under Contract No. DPT2006K-120470; National Research Foundation of Korea under Contract No. NRF-2022R1A2C1092335; National Science and Technology fund of Mongolia; National Science Research and Innovation Fund (NSRF) via the Program Management Unit for Human Resources \& Institutional Development, Research and Innovation of Thailand under Contracts Nos. B16F640076, B50G670107; Polish National Science Centre under Contract No. 2019/35/O/ST2/02907; Swedish Research Council under Contract No. 2019.04595; The Swedish Foundation for International Cooperation in Research and Higher Education under Contract No. CH2018-7756; U. S. Department of Energy under Contract No. DE-FG02-05ER41374.

\end{document}